\documentclass{article}
\usepackage{amssymb}
\usepackage{amsmath}

\input{tcilatex}

\begin{document}

\title{\textbf{The Two Envelope Problem:}\\
\textbf{a Paradox or Fallacious Reasoning?}}
\author{\textbf{Aris Spanos} \\
%EndAName
Department of Economics,\\
Virginia Tech, Blacksburg, VA 24061\\
\TEXTsymbol{<}aris@vt.edu\TEXTsymbol{>}\medskip }
\date{January 2013}
\maketitle

\begin{abstract}
The primary objective of this note is to revisit the two envelope problem
and propose a simple resolution. It is argued that the paradox arises from
the ambiguity associated with the money content \$$x$ of the chosen
envelope. When $X\mathbf{=}x$ is observed it is not know which one of the
two events, $X\mathbf{=}\theta $ or $X\mathbf{=}2\theta $, has occurred.
Moreover, the money in the other envelope $Y$ is not independent of $X$;
when one contains $\theta $ the other contains $2\theta .$ By taking these
important features of the problem into account, the paradox disappears.
\end{abstract}

\section{Introduction}

Consider two \textit{indistinguishable} envelopes that contain \$$\theta >0$
and \$$2\theta .$ Player 1 chooses one of the envelopes at random and
observes its content $X\mathbf{=}x\footnote{%
In certain variants of the paradox the player does not see $x$, but that
makes no difference to the following discussion.}$. The player is given the
choice to either keep \$$x$, or exchange it with the contents $Y$ of the
other envelope. What should player 1 do?

The traditional account is that `rational' reasoning by player 1 will
evaluate the expected value of $Y,$ defined in terms of $x$:%
\[
\begin{tabular}{ll}
$Y\mathbf{=}\frac{x}{2},$ & $\mathbb{P}\left( Y\mathbf{=}\frac{x}{2}\right) =%
\frac{1}{2},\medskip $ \\ 
$Y\mathbf{=}2x,$ & $\mathbb{P}\left( Y\mathbf{=}2x\right) =\frac{1}{2}.$%
\end{tabular}%
\]%
Hence, player's 1 expected winnings by trading envelopes will be:%
\begin{equation}
\begin{array}{c}
E(Y)=\left( \frac{x}{2}\right) \mathbb{P}\left( Y\mathbf{=}\frac{x}{2}%
\right) +2x\mathbb{P}\left( Y\mathbf{=}2x\right) =\left( \frac{x}{2}\right)
\left( \frac{1}{2}\right) +2x\left( \frac{1}{2}\right) =\frac{5}{4}x>x.%
\end{array}
\label{f}
\end{equation}%
This suggests that it will be rational for player 1 \textit{to always
exchange his envelope}, whatever the value $x$. This reasoning is clearly
fallacious, but the difficulty is to pinpoint the source of the problem; see
Nalebuff (1989), Broome (1995), Chalmers (2002), Clark and Shackel (2000),
Dietrich and List (2004), Falk and Nickerson (2009) inter alia.

It is argued that the paradox arises because of the ambiguity of the event $X%
\mathbf{=}x$ stemming from the fact that the observed value $x$ stands for
two different but \textit{unknown} values $\theta $ or $2\theta .$ That is,
when $X\mathbf{=}x$ is observed one does \textit{not} know which event $X%
\mathbf{=}\theta $ or $X\mathbf{=}2\theta \ $has occurred. Moreover, the
traditional account uses the marginal distribution of $Y$ expressed in terms
of the equivocal event $X\mathbf{=}x$, when in fact the random variables $X$
and $Y$ are dependent; when one envelope contains $\theta $ the other
contains $2\theta .$ When these features are taken into account the paradox
vanishes.

\section{A paradox or fallacious reasoning?}

The first issue to reconsider is the nature of the random variable $X$
denoting the money in the envelope initially chosen by Player 1. The player
observes its content $X\mathbf{=}x,$ but does not know is whether $x$
represents $\theta $ or $2\theta .$ Hence, the random variable $X$ is, in
effect, \textit{latent}:%
\[
X\mathbf{=}\left\{ 
\begin{array}{cc}
\theta & \text{for }x\mathbf{=}\theta \\ 
2\theta & \text{for }x\mathbf{=}2\theta%
\end{array}%
\right. ,\ 
\begin{array}{l}
\mathbb{P}(X\mathbf{=}\theta )\mathbf{=}.5 \\ 
\mathbb{P}(X\mathbf{=}2\theta )\mathbf{=}.5%
\end{array}%
\]%
The probability $.5$ arises from the fact that the two envelopes are
indistinguishable. In light of the fact that when one of the envelopes
contains \$$\theta $ the other must contain \$$2\theta $, the relevant
distribution is the joint distribution of $X$ and $Y,$ given in table 1. It
is important to note is that the support of both random variables, $R_{X}%
\mathbf{=}\{x$: $f(x)>0\}$ and $R_{Y}\mathbf{=}\{y$: $f(y)>0\},$ depends on
the unknown parameter $\theta $, rendering them \textit{non-regular}; see
Cox and Hinkley (1974). 
\begin{equation}
\begin{tabular}{|c||c|c||c|}
\hline
\multicolumn{4}{|c|}{\textbf{Table 1}} \\ \hline\hline
$X\ \backslash \ Y$ & $\theta $ & $2\theta $ & $\ f(x)$ \\ \hline\hline
$\theta $ & $0$ & $.5$ & $\underset{\quad }{\overset{\quad }{.5}}$ \\ \hline
$2\theta $ & $.5$ & $0$ & $\underset{\quad }{\overset{\quad }{.5}}$ \\ 
\hline\hline
$\ f(y)$ & $\underset{\quad }{\overset{\quad }{.5}}$ & $.5$ & $1$ \\ \hline
\end{tabular}%
\end{equation}%
Not surprisingly, $f(x,y)\neq f(x)\cdot f(y),$ for all $\left( x,y\right) ,$
and thus $X$ and $Y$ are \textit{not} independent, but the problem is now
symmetric with respect to both random variables. Indeed, the expected
winnings from either envelope are identical:%
\begin{equation}
\begin{array}{cc}
E(X) & =.5\theta +.5(2\theta )=1.5\theta ,\medskip \\ 
E(Y) & =.5\theta +.5(2\theta )=1.5\theta ,%
\end{array}
\label{E}
\end{equation}%
rendering player 1 indifferent between retaining \$$x$ or exchanging
envelopes. This result shows that whether player 1 should exchange depends
crucially on the relationship between the observed value $x$ and $\theta :$

(i) for $x\mathbf{=}2\theta ,$ $E(Y)\mathbf{=}\frac{3}{4}x<x,$ and thus
player 1 should \textit{not} exchange, but

(ii) for $x\mathbf{=}\theta ,$ $E(Y)\mathbf{=}\frac{3}{2}x>x,$ and player 1
should exchange.\newline
The problem, however, is that observing $X\mathbf{=}x$ is inadequate to make
an informed decision whether to exchange or not. This resolves the paradox!

\section{Conditioning on latent variables}

A more circuitous\ but illuminating way to reach the same conclusion is to
treat both random variables as latent and deal with the ambiguity of the
event $X\mathbf{=}x$ using the conditional expectation $E(Y\mathbf{\mid }%
\sigma (X)),$ where $\sigma (X)\mathbf{=}\{S,\varnothing ,X\mathbf{=}\theta
,X\mathbf{=}2\theta \}$ denotes the sigma-field generated by $X.$ Since $%
\sigma (X)\mathbf{\subset }\mathcal{F},$ conditioning on $\sigma (X)$ simply
acknowledges the possible events generated by $X$ via restricting the
universal $\mathcal{F}$ related to the original probability space $\left( S,%
\mathcal{F},\mathbb{P}(.)\right) ,$ upon which both random variables $\left(
X,Y\right) $ have been defined. Formally, conditioning on $\sigma (X)$
constitutes a restriction because:%
\[
E(Y\mathbf{\mid }\mathcal{F})\mathbf{=}Y\text{ but }E(Y\mathbf{\mid }\sigma
(X))\mathbf{=}g(X)\neq Y.
\]%
Moreover, the random variable $E(Y$\textbf{$\mid $}$\sigma (X))$ does not
depend on the \textit{particular values} $x$ of $X$ because for any Borel
function $h(.)$ which keeps those values distinct, i.e. for two different
values of $X,$ say $x_{1}\neq x_{2},$ $h(x_{1})\neq h(x_{2})$ (Renyi, 1970,
p. 259):%
\[
E(Y\mathbf{\mid }\sigma (X))\mathbf{=}E(Y\mathbf{\mid }\sigma (h(X))),\text{
since }\sigma (X)\mathbf{=}\sigma (h(X)).
\]

To evaluate $E(Y\mathbf{\mid }\sigma (X))$ one needs both conditional
distributions:%
\[
f(Y\mathbf{\mid }X\mathbf{=}\theta )\mathbf{=}\left\{ 
\begin{array}{cc}
\frac{f(y\mathbf{=}2\theta ,x\mathbf{=}\theta )}{f(x\mathbf{=}\theta )}%
\mathbf{=}\frac{.5}{.5}\mathbf{=}1, & \text{ for }Y\mathbf{=}2\theta \\ 
\frac{f(y\mathbf{=}\theta ,x\mathbf{=}\theta )}{f(x\mathbf{=}\theta )}%
\mathbf{=}\frac{0}{.5}\mathbf{=}0, & \text{for }Y\mathbf{=}\theta%
\end{array}%
\right.
\]%
\[
f(Y\mathbf{\mid }X\mathbf{=}2\theta )\mathbf{=}\left\{ 
\begin{array}{cc}
\frac{f(y\mathbf{=}2\theta ,x\mathbf{=}2\theta )}{f(x\mathbf{=}2\theta )}%
\mathbf{=}\frac{0}{.5}\mathbf{=}0, & \text{ for }Y\mathbf{=}2\theta \\ 
\frac{f(y\mathbf{=}\theta ,x\mathbf{=}2\theta )}{f(x\mathbf{=}2\theta )}%
\mathbf{=}\frac{.5}{.5}\mathbf{=}1, & \text{for }Y\mathbf{=}\theta%
\end{array}%
\right.
\]%
Hence, $E(Y\mathbf{\mid }\sigma (X))$ defines a random variable of the form:%
\begin{equation}
E(Y\mathbf{\mid }\sigma (X))\mathbf{=}\left[ 2\theta +0\cdot \theta \right] 
\mathbb{I}_{\{x\mathbf{=}\theta \}}+\left[ \theta +0\cdot 2\theta \right] 
\mathbb{I}_{\{x\mathbf{=}2\theta \}}=2\theta \mathbb{I}_{\{x\mathbf{=}\theta
\}}+\theta \mathbb{I}_{\{x\mathbf{=}2\theta \}},  \label{ce}
\end{equation}%
where $\mathbb{I}_{\{x\mathbf{=}\theta \}}$ is the indicator function. To
derive the expected winnings of exchanging envelopes one needs $E(Y)$ which
can be derived from (\ref{ce}) using the law iterated expectations
(Williams, 1991): 
\begin{equation}
\begin{array}{c}
E(Y)\mathbf{=}\underset{X}{E}\{E(Y\mathbf{\mid }\sigma (X))\}=2\theta
(.5)+\theta (.5)=1.5\theta ,%
\end{array}
\label{lie}
\end{equation}%
which coincides with the result in (\ref{E}).

\section{The fallacy and the induced distribution of $\protect\theta $}

One might object to the reasoning giving rise to the evaluation of $E(Y%
\mathbf{\mid }\sigma (X))$ in (\ref{lie}) by claiming that one \textit{can}
attach probabilities to $x\mathbf{=}\theta $ and $x\mathbf{=}2\theta .$
Indeed, this has been the basis of several Bayesian solutions to this
paradox that often revolve around the conditional probabilities:\vspace*{%
-0.1in}%
\[
\begin{array}{ccc}
\mathbb{P}(X\mathbf{=}x\mathbf{\mid }\theta \mathbf{=}x), &  & \mathbb{P}(X%
\mathbf{=}x\mathbf{\mid }\theta \mathbf{=}\frac{x}{2}),%
\end{array}%
\]%
stemming from some form of prior information; see Christensen and Utts
(1992) and Lindley (2006). This move, however, invokes the potential
ambiguity between the event $X\mathbf{=}\theta $ and the value assignment $%
\theta \mathbf{=}x;$ the former is a legitimate frequentist event, but the
latter constitutes an event only in the context of Bayesian inference. This
ambiguity can inadvertently give rise to creating an induced distribution
for $\theta $. This can easily arise when an overlap between the parameter
and sample spaces has been created by a non-regular distribution. This
overlap could misleadingly be used to derive the `induced' distribution of $%
\theta $ from that of $X:$%
\begin{equation}
\begin{tabular}{|l||l|}
\hline
\multicolumn{2}{|l|}{\textbf{Table 2}} \\ \hline\hline
$x$ & $f(x)$ \\ \hline\hline
$\overset{\quad }{\theta }$ & $\underset{\quad }{\overset{\quad }{.5}}$ \\ 
\hline
$\overset{\quad }{2\theta }$ & $\underset{\quad }{\overset{\quad }{.5}}$ \\ 
\hline
\end{tabular}%
\Rightarrow 
\begin{tabular}{|l||l|}
\hline
\multicolumn{2}{|l|}{\textbf{Table 3}} \\ \hline\hline
$\theta $ & $p(\theta )$ \\ \hline\hline
$\overset{\quad }{x}$ & $\underset{\quad }{\overset{\quad }{.5}}$ \\ \hline
$\overset{\quad }{\frac{x}{2}}$ & $\underset{\quad }{\overset{\quad }{.5}}$
\\ \hline
\end{tabular}
\label{2d}
\end{equation}%
and then (inadvertently) proceed to use $p(\theta )$ in place of $f(x)$.

To demonstrate how conflating $X\mathbf{=}\theta $ and $X\mathbf{=}2\theta $
with $x\mathbf{=}\theta $ and $x\mathbf{=}2\theta $ can lead to the
fallacious result (\ref{f}), consider replacing $f(x)$ (table 2) with $%
p(\theta )$ (table 3) in (\ref{ce}). This replacement yields:\vspace*{-0.1in}%
\begin{equation}
\begin{array}{c}
E^{\ddag }(Y\mathbf{\mid }\sigma (X))=2x\mathbb{I}_{\{x\mathbf{=}\theta \}}+%
\frac{x}{2}\mathbb{I}_{\{x\mathbf{=}2\theta \}}%
\end{array}%
\Rightarrow
\end{equation}%
\[
\begin{array}{c}
E^{\ddag }(Y)\mathbf{=}\underset{x}{E}\{E^{\ddag }(Y\mathbf{\mid }\sigma
(X))\}\mathbf{=}2x(\frac{1}{2})+\frac{x}{2}(\frac{1}{2})\mathbf{=}\frac{5}{4}%
x,%
\end{array}%
\]%
which coincides with the fallacious expected value in (\ref{f}).

This confusion can be seen in the evaluation of the likelihood function:%
\begin{equation}
\begin{array}{rl}
L(\theta \mathbf{=}x) & \mathbf{=}\mathbb{P}(X\mathbf{=}x\mathbf{\mid }%
\theta \mathbf{=}x)\mathbf{=}\mathbb{P}(X\mathbf{=}\theta \mathbf{\mid }%
\theta \mathbf{=}x)\mathbf{=}.5,\medskip \\ 
L(\theta \mathbf{=}\frac{x}{2}) & \mathbf{=}\mathbb{P}(X\mathbf{=}x\mathbf{%
\mid }\theta \mathbf{=}\frac{x}{2})\mathbf{=}\mathbb{P}(X\mathbf{=}2\theta 
\mathbf{\mid }\theta \mathbf{=}\frac{x}{2})\mathbf{=}.5,%
\end{array}
\label{L}
\end{equation}%
given in Pawitan (2001), p. 27.

\section{Conclusion}

The key conclusion is that the two envelope (exchange) paradox stems
primarily from the ambiguity associated with the money content \$$x$ of the
chosen envelope. When the event $X\mathbf{=}x$ is observed one does not know
which of the two different events, $X\mathbf{=}\theta $ or $X\mathbf{=}%
2\theta ,$ has occurred. Moreover, the money content of the other envelope $%
Y $ is dependent on $X;$ if one contains \$$\theta $ the other contains \$$%
2\theta .$ The appropriate way to deal with these features of the problem is
to use treat both random variables as \textit{latent} and derive $E(Y)$
either directly or via $E(Y\mathbf{\mid }\sigma (X))$.

Taking these features into account resolves the paradox because:%
\[
\begin{array}{c}
E(Y)\mathbf{=}\underset{X}{E}\{E(Y\mathbf{\mid }\sigma (X))\}\mathbf{=}%
1.5\theta \neq E^{\ddag }(Y)\mathbf{=}\left( \frac{x}{2}\right) \mathbb{P}%
\left( Y\mathbf{=}\frac{x}{2}\right) +2x\mathbb{P}\left( Y\mathbf{=}%
2x\right) \mathbf{=}1.25x.%
\end{array}%
\]%
The result $E(Y)\mathbf{=}1.5\theta $ indicates that the optimal strategy
for player 1 depends crucially on whether $x\mathbf{=}\theta $ or $x\mathbf{=%
}2\theta .$ Without the latter information, player 1 is indifferent between
the two envelopes since $E(Y)\mathbf{=}E(X)=1.5\theta $.

\end{document}